\documentclass{article}
\usepackage{LaThuileFPSpro}
\begin{document}
\newcommand{\mc}{\multicolumn}
\newcommand{\bce}{\begin{center}}
\newcommand{\ece}{\end{center}}
\newcommand{\beq}{\begin{equation}}
\newcommand{\eeq}{\end{equation}}
\newcommand{\bea}{\begin{eqnarray}}

\newcommand{\eea}{\end{eqnarray}}
\newcommand{\cont}{\nonumber\eea\bea}
\newcommand{\cl}[1]{\begin{center} {#1} \end{center}}
\newcommand{\ba}{\begin{array}}
\newcommand{\ea}{\end{array}}
%\newcommand{\arr}{\bea}

% -------------- MATH def -------------------------------
\newcommand{\ab}{{\alpha\beta}}
\newcommand{\cd}{{\gamma\delta}}
\newcommand{\dc}{{\delta\gamma}}
\newcommand{\ac}{{\alpha\gamma}}
\newcommand{\bd}{{\beta\delta}}
\newcommand{\abc}{{\alpha\beta\gamma}}
\newcommand{\eps}{{\epsilon}}
\newcommand{\lam}{{\lambda}}
\newcommand{\mn}{{\mu\nu}}
\newcommand{\mpnp}{{\mu'\nu'}}
\newcommand{\Amuu}{{A_{\mu}}}
\newcommand{\Amuo}{{A^{\mu}}}
\newcommand{\Vmuu}{{V_{\mu}}}
\newcommand{\Vmuo}{{V^{\mu}}}
\newcommand{\Anuu}{{A_{\nu}}}
\newcommand{\Anuo}{{A^{\nu}}}
\newcommand{\Vnuu}{{V_{\nu}}}
\newcommand{\Vnuo}{{V^{\nu}}}
\newcommand{\Fmnu}{{F_{\mu\nu}}}
\newcommand{\Fmno}{{F^{\mu\nu}}}

\newcommand{\abcd}{{\alpha\beta\gamma\delta}}

% Boldmath definitions

\newcommand{\bsigma}{\mbox{\boldmath $\sigma$}}
\newcommand{\btau}{\mbox{\boldmath $\tau$}}
\newcommand{\brho}{\mbox{\boldmath $\rho$}}
\newcommand{\bpipi}{\mbox{\boldmath $\pi\pi$}}
\newcommand{\bss}{\bsigma\!\cdot\!\bsigma}
\newcommand{\btt}{\btau\!\cdot\!\btau}
\newcommand{\bnabla}{\mbox{\boldmath $\nabla$}}
\newcommand{\bphi}{\mbox{\boldmath $\tau$}}
\newcommand{\bvarphi}{\mbox{\boldmath $\rho$}}
\newcommand{\bDelta}{\mbox{\boldmath $\Delta$}}
\newcommand{\bpsi}{\mbox{\boldmath $\psi$}}
\newcommand{\bPsi}{\mbox{\boldmath $\Psi$}}
\newcommand{\bPhi}{\mbox{\boldmath $\Phi$}}
\newcommand{\bnab}{\mbox{\boldmath $\nabla$}}
\newcommand{\bpi}{\mbox{\boldmath $\pi$}}
\newcommand{\btheta}{\mbox{\boldmath $\theta$}}
\newcommand{\bkappa}{\mbox{\boldmath $\kappa$}}

\newcommand{\bA}{{\bf A}}
\newcommand{\bfe}{{\bf e}}
\newcommand{\bb}{{\bf b}}
\newcommand{\br}{{\bf r}}
\newcommand{\bj}{{\bf j}}
\newcommand{\bk}{{\bf k}}
\newcommand{\bl}{{\bf l}}
\newcommand{\bL}{{\bf L}}
\newcommand{\bM}{{\bf M}}
\newcommand{\bp}{{\bf p}}
\newcommand{\bq}{{\bf q}}
\newcommand{\bR}{{\bf R}}
\newcommand{\bs}{{\bf s}}
\newcommand{\bS}{{\bf S}}
\newcommand{\bT}{{\bf T}}
\newcommand{\bv}{{\bf v}}
\newcommand{\bV}{{\bf V}}
\newcommand{\bx}{{\bf x}}
\newcommand{\fph}{${\cal F}$}
\newcommand{\aph}{${\cal A}$}
\newcommand{\dph}{${\cal D}$}
\newcommand{\fpi}{f_\pi}
\newcommand{\mpi}{m_\pi}
\newcommand{\Tr}{{\mbox{\rm Tr}}}
\def\Qb{\overline{Q}}
\newcommand{\delu}{\partial_{\mu}}
\newcommand{\delo}{\partial^{\mu}}
%\newcommand{\half}{{1\over 2}}
%\newcommand{\quart}{{1\over 4}}
%
%
% ------------------ arrow mod ---------------------
\newcommand{\up}{\!\uparrow}
\newcommand{\upup}{\uparrow\uparrow}
\newcommand{\updo}{\uparrow\downarrow}
\newcommand{\uu}{$\uparrow\uparrow$}
\newcommand{\ud}{$\uparrow\downarrow$}
\newcommand{\auu}{$a^{\uparrow\uparrow}$}
\newcommand{\aud}{$a^{\uparrow\downarrow}$}
\newcommand{\pu}{p\!\uparrow}

% ------------------------------------------------------
\newcommand{\qp}{quasiparticle}
\newcommand{\sa}{scattering amplitude}
\newcommand{\ph}{particle-hole}
\newcommand{\qcd}{{\it QCD}}
\newcommand{\integ}{\int\!d}
\newcommand{\ie}{{\sl i.e.~}}
\newcommand{\etal}{{\sl et al.~}}
\newcommand{\etc}{{\sl etc.~}}
\newcommand{\rhs}{{\sl rhs~}}
\newcommand{\lhs}{{\sl lhs~}}
\newcommand{\eg}{{\sl e.g.~}}
\newcommand{\ef}{\epsilon_F}
\newcommand{\sigt}{d^2\sigma/d\Omega dE}
\newcommand{\sige}{{d^2\sigma\over d\Omega dE}}
% ----------------------- ------------------------------
\newcommand{\rpaeq}{\beq
\left ( \begin{array}{cc}
A&B\\
-B^*&-A^*\end{array}\right )
\left ( \begin{array}{c}
X^{(\kappa})\\Y^{(\kappa)}\end{array}\right )=E_\kappa
\left ( \begin{array}{c}
X^{(\kappa})\\Y^{(\kappa)}\end{array}\right )
\eeq}
\newcommand{\ket}[1]{| {#1} \rangle}
\newcommand{\bra}[1]{\langle {#1} |}
\newcommand{\ave}[1]{\langle {#1} \rangle}

%\newcounter{f1}
%\newcounter{f2}
%\renewcommand{\theequation}{\thesubsection.\arabic{equation}}
%\renewcommand{\thetable}{\thesection.\arabic{table}}
\newcommand{\singlespace}{
    \renewcommand{\baselinestretch}{1}\large\normalsize}
\newcommand{\doublespace}{
    \renewcommand{\baselinestretch}{1.6}\large\normalsize}
\newcommand{\bftau}{\mbox{\boldmath $\tau$}}
\newcommand{\bfalpha}{\mbox{\boldmath $\alpha$}}
\newcommand{\bfgamma}{\mbox{\boldmath $\gamma$}}
\newcommand{\bfxi}{\mbox{\boldmath $\xi$}}
\newcommand{\bfbeta}{\mbox{\boldmath $\beta$}}
\newcommand{\bfeta}{\mbox{\boldmath $\eta$}}
\newcommand{\bfpi}{\mbox{\boldmath $\pi$}}
\newcommand{\bfphi}{\mbox{\boldmath $\phi$}}
\newcommand{\bfR}{\mbox{\boldmath ${\cal R}$}}
\newcommand{\bfL}{\mbox{\boldmath ${\cal L}$}}
\newcommand{\bfM}{\mbox{\boldmath ${\cal M}$}}
\def\dblint{\mathop{\rlap{\hbox{$\displaystyle\!\int\!\!\!\!\!\int$}}
    \hbox{$\bigcirc$}}}
\def\ut#1{$\underline{\smash{\vphantom{y}\hbox{#1}}}$}

\def\UNITY{{\bf 1\! |}}
\def\Pom{{\bf I\!P}}
\def\lsim{\mathrel{\rlap{\lower4pt\hbox{\hskip1pt$\sim$}}
    \raise1pt\hbox{$<$}}}         %less than or approx. symbol
\def\gsim{\mathrel{\rlap{\lower4pt\hbox{\hskip1pt$\sim$}}
    \raise1pt\hbox{$>$}}}         %greater than or approx. symbol
\def\beq{\begin{equation}}
\def\eeq{\end{equation}}
\def\bea{\begin{eqnarray}}
\def\eea{\end{eqnarray}}
\title{DECORRELATION OF FORWARD DIJETS IN DIS OFF NUCLEI
\footnote{Talk given at Les Rencontres de Physique de la Valle d'Aoste,
La Thuile, Aoste Valley, Italy  9-15 March 2003}} 

\author{
  V.R. Zoller        \\
  {\em ITEP, Moscow, Russia} \\
}
\maketitle

\baselineskip=11.6pt

\begin{abstract}
Based on the color dipole QCD approach  we discuss the 
multiple scattering mechanism (MSM) of suppression   
of back-to-back azimuthal correlations
of high $p_{T}$ forward dijets in DIS off nuclei.
We quantify the effect in terms of acoplanarity/decorrelation 
 momentum of jets. For hard
dijets the decorrelation  momentum is found to be of the order
of the nuclear saturation momentum $Q_A$. Minijets with the
transverse momentum below the saturation scale are proved to be
completely decorrelated.  The analysis of DIS  indicates that 
in heavy ion collisions the contribution 
of MSM to the reactions dynamics
can be of the order of magnitude of the decorrelation 
effect observed at RHIC.

\end{abstract}
\newpage
\section{Introduction}

Recent observation  of gradual 
disappearance of azimuthal 
back-to-back correlations of high $p_T$ particles with centrality of $Au\,Au$
collisions at  RHIC \cite{white} is presently viewed as 
a consequence of the jet energy loss
in hot quark-gluon plasma produced in central collisions. 
Another possible explanation is that 
initial and final
state interactions associated with multiple parton scatterings
on hard stage of the process produce uncorrelated monojets. 

To get an idea of the size  of the multiple scatterings (MS) effect at RHIC
we start with the QCD description of the breakup of 
photons into forward dijets in small-$x$ deep inelastic scattering (DIS)
 off nuclei
in the saturation regime \cite{NSZZ}. 
In \cite{NSZZ} we reported a 
derivation of the general formula for the two-body transverse momentum
distribution. 
Our formalism, based on the technique \cite{NZ91,NPZcharm,LPM}, includes 
consistently the diffractive attenuation of color dipoles and
effects of transitions between
different color states of the  $q\bar{q}$-pair propagating through
the nucleus. 
Opacity of nuclei brings in a new  scale $Q_A$
which separates the regimes of opaque nuclei and weak attenuation
\cite{Mueller,Mueller1,McLerran,Saturation}. For hard
dijets the decorrelation  momentum is found to be of the order
of the nuclear saturation momentum $Q_A$ \cite{NSZZ}.
For  parton momenta below the saturation scale
$Q_A$
the evolution of sea from gluons was shown to be dominated
by the anti-collinear, anti-DGLAP splitting \cite{Saturation}. 
As a result, minijets with the
transverse momentum below the saturation scale are proved to be
completely decorrelated \cite{NSZZ}. 

Turning back to the RHIC observation  we argue
that MS mechanism may contribute  substantially to the effect of vanishing
 of back-to-back azimuthal correlations of high
$p_T$ hadrons in central nucleus-nucleus collisions.

\section{Breakup of photons into hard dijets on nuclear targets}
 We consider DIS at 
$x\lsim x_A = 1/R_A m_N \ll 1$ which is dominated by interactions 
of $q\bar{q}$ Fock states of the photon and make use of the conventional 
approximation of two t-channel gluons
in DIS off free nucleons \cite{GunSop}. The two-gluon
exchange approximation amounts to neglecting diffractive DIS off free 
nucleons  which
is justified by a small fraction of diffractive DIS, $\eta_D \ll 1$
\cite{GNZ95}. We sum  unitarity cuts of the
forward Compton scattering amplitude which 
 describe the transition from the color-neutral $q\bar{q}$
dipole to the color-octet $q\bar{q}$ pair.  It should be emphasized that 
we are interested in the unitarity cuts  which correspond to 
the genuine inelastic DIS with color
excitation of the target nucleus. 

For $x\lsim x_A$ the
propagation of the $q\bar{q}$ pair inside nucleus can be treated 
in the straight-path approximation.
Let $\bb_+$ and $\bb_-$ be the impact parameters of the quark
and antiquark, respectively, and  $S_A(\bb_+,\bb_-)$   be the
S-matrix for interaction of the $q\bar{q}$ pair with the nucleus. 
 Regarding
the color states $|c\rangle=|1\rangle,\,|8\rangle$ of the $q\bar{q}$ pair, 
we sum over all octet
and singlet states. Then the 2-jet inclusive
spectrum is calculated in terms of the 2-body density matrix as
\bea
&&{d\sigma_{in} \over dz d^2\bp_+ d^2\bp_-} =
{1\over (2\pi)^4} \int d^2 \bb_+' d^2\bb_-' d^2\bb_+ d^2\bb_- \nonumber\\
&&\times \exp[-i\bp_+(\bb_+ -\bb_+')-i\bp_-(\bb_- -
\bb_-')]\Psi^*
\Psi\nonumber\\
&&\times \Bigl\{ \sum_{A^*} \sum_{c}  \langle
1;A|S_A^*|A^*;c\rangle \langle
c;A^*|S_A|A;1\rangle  \nonumber\\
&& -
\langle 1;A|S_A^*
|A;1\rangle
\langle 1;A|S_A|A;1\rangle \Bigr\}\, .
\label{eq:3.1}
\eea
 In (\ref{eq:3.1}) $\Psi=\Psi(Q^2,z,\bb_+ -\bb_-)$ stands for
 the  wave function of the
$q\bar{q}$ Fock state of the photon with virtuality $Q^2$ and the 
photon light-cone momentum fraction $z$ carried by the quark.
 Notice, that the calculation of the 
2-body density matrix  enter four straight-path trajectories
$\delta^{(2)}(\bb_{\pm})\,,\delta^{(2)}(\bb'_{\pm})$   
and $S_A$ and $S_A^*$ describe the propagation of
two quark-antiquarks pairs, $q\bar{q}$ and $q'\bar{q}'$, 
inside a nucleus. In the
integrand of  (\ref{eq:3.1}) we subtracted the coherent 
diffractive component
of the final state. 
  
Upon the application of closure to sum over nuclear final states
$A^*$ the integrand of (\ref{eq:3.1}) can be considered as an
intranuclear evolution operator for the 2-body density matrix
(for the related
discussion see refs. \cite{BGZpositronium},\cite{NSZdist}) 
 \bea \sum_{A^*}
\sum_{c} \langle A| \Bigl\{ \langle 1|
S_A^*|c\rangle \Bigr\} |A^* \rangle \langle
A^*| \Bigl\{ \langle c| S_A|1\rangle \Bigr\}
|A\rangle =\nonumber\\
=\langle A| \left\{ \sum_{c} \langle 1|
S_A^*|c\rangle \langle
c|S_A|1\rangle\right\} |A\rangle \, .
\label{eq:3.2}
\eea 
Let the eikonal for the quark-nucleon and
antiquark-nucleon QCD gluon exchange interaction be $T^a_{+}
\chi(\bb)$ and  $T^a_{-} \chi(\bb)$, where  $T^{a}_+$ and
$T^{a}_{-}$ are the $SU(N_c)$ generators for the quark and
antiquarks states, respectively. The vertex $V_a$ for excitation
of the nucleon $g^a N \to N^*_a$ into color octet state is so
normalized that after application of closure the vertex $g^a g^b
NN$  is $\delta_{ab}$. Then, to the
two-gluon exchange approximation, the $S$-matrix of the
$(q\bar{q})$-nucleon interaction equals 
\bea 
S_N(\bb_+,\bb_-)  = 1
+ i[T^a_{+} \chi(\bb_+)+ T^a_{-} \chi(\bb_-)]V_a - {1\over 2}
[T^a_{+} \chi(\bb_+)+ T^a_{-} \chi(\bb_-)]^2 \, .
\label{eq:3.3}
\eea
The profile function for interaction of the $q\bar{q}$ dipole
with a nucleon is 
$\Gamma(\bb_+,\bb_-)= 1 - S_N(\bb_+,\bb_-)$.
For a color-singlet dipole 
$(T^a_{+}+ T^a_{-})^2=0$ and 
the dipole cross section for interaction of the color-singlet
$q\bar{q}$ dipole with the nucleon equals
\bea
\sigma(\bb_+-\bb_-) = 2\int d^2\bb_{+} \langle N|\Gamma(\bb_+,\bb_-)
|N\rangle \nonumber\\ 
=
{N_c^2 -1 \over 2 N_c} \int
d^2\bb_+ [\chi(\bb_+)-\chi(\bb_-)]^2\, . 
\label{eq:3.4} \eea
The nuclear $S$-matrix of the straight-path approximation is
$$
S_A(\bb_+,\bb_-) = \prod _{j=1}^A S_N(\bb_+ -\bb_j,\bb_- - \bb_j)\, ,
$$
where the ordering along the longitudinal path is understood.
We evaluate the nuclear expectation value in (\ref{eq:3.2})
in the standard dilute gas approximation.
To the two-gluon exchange approximation, per each and every nucleon
$N_j$ only the terms quadratic
in $\chi(\bb_j)$ must be kept in the evaluation of the
single-nucleon matrix element
$
\langle N_j|S_N^*S_N|N_j \rangle
$
 which enters the calculation of
$S_A^*S_A$. Following the technique developed in
\cite{NPZcharm,LPM} we can reduce the calculation of the evolution
operator for the 2-body density matrix (\ref{eq:3.2}) to the
evaluation of the $S$-matrix $S_{4A}(\bb_+,\bb_-,\bb_+',\bb_-')$
for the scattering of a fictitious 4-parton state composed of the
two quark-antiquark pairs in the overall color-singlet state.
Because $(T^{a}_+)^*= -T^{a}_{-}$, within the two-gluon exchange
approximation the quarks entering the complex-conjugate $S_A^*$ in
(\ref{eq:3.2}) can be viewed as antiquarks, so that 
\bea
\sum_{c} \langle 1| S_A^*|c\rangle \langle
c|S_A|1\rangle 
=\sum_{c^{\prime} c} \delta_{c^{\prime} c}
\langle c^{\prime}c|S_{4A}|11\rangle \, ,
\label{eq:3.5}
\eea
where $S_{4A}(\bb_+',\bb_-',\bb_+,\bb_-)$ is an
S-matrix for the propagation of the two quark-antiquark pairs in the
overall singlet state. While the first $q\bar{q}$ pair is formed
by the initial quark $q$ and antiquark $\bar{q}$ at impact parameters
$\bb_{+}$ and $\bb_-$, respectively, in the second $q'\bar{q}'$ pair 
the quark $q'$ propagates at an impact parameter $\bb_-'$ and the
antiquark $\bar{q}'$ at an impact parameter $\bb_+'$.

If $\sigma_4(\bb_+',\bb_-',\bb_+,\bb_-)$ is the color-dipole cross
section operator for the 4-body state, then the 
evaluation of the nuclear expectation value for a dilute gas
nucleus in the standard approximation of
neglecting the size of color dipoles compared to a radius of heavy
nucleus gives
\cite{Glauber}
 \bea S_{4A}(\bb_+',\bb_-',\bb_+,\bb_-)=\exp\{-
{1\over 2}\sigma_{4}(\bb_+',\bb_-',\bb_+,\bb_-)T(\bb)\}\, ,
\label{eq:3.9} 
\eea 
where 
$ T(\bb)=\int db_z n_{A}(b_z, \bb)$ 
is the optical thickness of a nucleus at an
impact parameter 
$
\bb = {1\over 4}( \bb_+ + \bb_+' + \bb_- + \bb_-')\, ,
$
 and $ n_{A}(b_z, \bb)$
is nuclear matter density with the normalization 
$\int d^2\bb T(\bb) = A$ .  The single-nucleon $S$-matrix (\ref{eq:3.3})
 contains
transitions from the color-singlet to the both color-singlet and
color-octet $q\bar{q}$ pairs. However, only  color-singlet
operators contribute to $\langle N_j|S_N^* S_N|N_j \rangle$, and hence
the matrix
$\sigma_4$ only includes transitions
between the $|11\rangle$ and $|88\rangle$ color-singlet 4-parton
states, the $|18\rangle$ states are not allowed.

After some color
algebra, we find matrix elements
$\sigma_{11}=\langle 11|\sigma_4|11\rangle$, $\sigma_{18}=\sigma_{81}=
\langle 11|\sigma_4|88\rangle  $ and 
$\sigma_{88}=\langle 88|\sigma_4|88\rangle$
 (details of the derivation are presented in
\cite{NSZZ}).
For  forward hard
jets with the momenta $\bp_{\pm}^2 \gg Q_A^2$,
which  are produced from
interactions with the target nucleus of small color dipoles
in the incident photon  the two eigenvalues of the operator $\sigma_{4}$  are
$$
\Sigma_2\approx \sigma_{11}
$$
and
$$
\Sigma_1\approx  \sigma_{88}.
$$
For small color-singlet dipoles
$
\Sigma_2\approx \sigma_{11} \approx 0
$
and the nuclear
distortion factor takes on a simple form 
\bea \sum_{A^*} \sum_{c}
\langle A|\langle 1|S_A^*|c\rangle \langle c|S_A|1\rangle |A\rangle  
- \langle 1;A|S_A^* |A;1\rangle
\langle 1;A|S_A|A;1\rangle  
\nonumber\\
=(\langle 11| + \sqrt{N_c^2 -1} \langle 88|) \exp\left\{-{1\over
2}\sigma_4 T(\bb)\right\}|11\rangle - 
\exp\left\{-{1\over2}\sigma_{11}
T(\bb)\right\}\nonumber\\
\approx
\sqrt{N_c^2 -1}\, {\sigma_{18} \over
\sigma_{88}} \left\{1- \exp\left[-{1\over 2}\sigma_{88} T(\bb)\right]\right\} 
\, .
\label{eq:3.18} \eea
The introduction of this nuclear distortion factor into (\ref{eq:3.1})
gives  the hard dijet inclusive cross section:
\bea
{d\sigma_{in} \over d^2\bb dz d^2\bp_+ d^2\bDelta}= 
T(\bb)\sum_{j=0}^{\infty}w_A(\bb,j) 
\int d^2\bkappa
f^{(j)}(\bDelta-\bkappa)
{d\sigma_{N} \over dz d^2\bp_+ d^2\bkappa }\, .
\label{eq:4.10}
\eea
where the acoplanarity/decorrelation momentum $\bDelta$ is
$$
\bDelta=\bp_+ + \bp_-,
$$
the probability of finding $j$ spatially overlapping nucleus in a 
Lorentz-contracted nucleus is
$$
w_A(\bb,j)=
{1 \over j!} {\gamma(j+1,2\lambda_c  \nu_A(\bb)) \over
2\lambda_c  \nu_A(\bb)},
$$
$$
\nu_A(\bb)={1\over 2}\alpha_S\sigma_0T(\bb),
$$
$$
\lambda_c=N_c^2/(N_c^2-1)
$$
and  $\gamma(j,x)= \int_{0}^x dy y^{j-1}\exp(-y)$ is an 
incomplete gamma-function. 
The function
\beq
f^{(j)}(\bkappa )= \int \prod_{i=1}^j
d^2\bkappa _{i} f(\bkappa _{i}) \delta(\bkappa -\sum_{i=1}^j
\bkappa _i) \,, ~~f^{(0)}(\bkappa)=\delta(\bkappa) 
\label{eq:4.8} 
\eeq 
in eq.(\ref{eq:4.10})
is a collective gluon field of $j$ overlapping nucleons 
introduced in \cite{NSSdijet}.

 The probabilistic form of a 
convolution  of the differential cross
section  on a free nucleon target with the manifestly positive 
defined distribution $f^{(j)}(\bkappa )$
in (\ref{eq:4.10})
can be
understood as follows. Hard jets
originate from small color dipoles. It is the quantum mechanical
 interference
that suppresses interaction with soft  gluons  
of  the small-sized 
color-singlet $q\bar{q}$ state.
However, the first inelastic
interaction inside a nucleus converts the $q\bar{q}$ pair into
the color-octet state, in which color charges of the quark and 
antiquark do not neutralize each other, rescatterings of the quark
and antiquark in the collective color field of intranuclear nucleons become
uncorrelated, and the broadening of the momentum
distribution with nuclear thickness follows a probabilistic
picture.

\section{Decorrelation of dijets in DIS off nuclei:
numerical estimates}

To quantify the azimuthal decorrelation  of two forward jets we find it convenient 
to introduce  the
mean transverse acoplanarity momentum squared 
$\langle \bDelta_{T}^2(\bb) \rangle$, where $\bDelta_{T}$ is
transverse to an axis of the jet with higher momentum. 
It is assumed that jets are hard, $|\bp_+| \gg Q_A$.
The
convolution property of the hard dijet cross section 
(\ref{eq:4.10}) suggests
\bea
\langle \bDelta_{T}^2(\bb) \rangle_{A} 
\approx   
\langle \bkappa_{T}^2(\bb) \rangle_A  +  
\langle \bDelta_{T}^2 \rangle_N  \, ,
\label{eq:6.1}
\eea
where $\langle \bDelta_{T}^2 \rangle_N$ refers to DIS
on a free nucleon, 
  and $ \langle\bkappa_{T}^2(\bb) \rangle_A $ is the nuclear
broadening term 
The sign $\approx$ in (\ref{eq:6.1}) 
reflects the kinematical limitations  on $\bp_-$ and 
$\bkappa$ in the practical 
evaluation of the acoplanarity distribution. In a typical final 
state  it is the harder jet with
larger transverse momentum which defines the jet axis and the
acoplanarity momentum $\bDelta$ will be defined in terms 
of components of the momentum of softer jet with respect to
that axis, for instance, see \cite{RHIC_STAR}. 
For the sake of definiteness, we present numerical 
estimates for the Gedanken experiment
in which we classify the event as a dijet if the quark 
and antiquark are produced in different hemispheres, i.e., if the
azimuthal angle $\phi$ between two jets exceeds $\pi/2$, the quark jet 
has fixed $|\bp_+|$ and  the antiquark jet has higher transverse momentum   
$|\bp_+| \lsim |\bp_-| \lsim 10 |\bp_+|$ 
(in the discussion of the experimental data
one often refers to the higher momentum jet as the trigger jet
and the softer jet as the away jet \cite{RHIC_STAR}).
The free-nucleon quantity $\langle \bDelta_{T}^2 \rangle_N$ 
can be estimated starting with
 the  
small-$\bDelta$ expansion for excitation of 
hard, $\bp_+^2 \gg \varepsilon^2=z(1-z)Q^2$,  light flavor
dijets from transverse photons   
\bea
{d\sigma_N \over dz d^2\bp_+ d^2\bDelta} & \approx &
{1\over \pi}  e_f^2 \alpha_{em}\alpha_S(\bp_+^2)\left[z^2 + (1-z)^2\right]
\nonumber\\
& \times & {1\over \Delta^4} 
\cdot{\partial G(x,\bDelta^2)\over \partial \log \bDelta^2}\cdot 
{\bDelta^2 \over (\varepsilon^2 +\bp_+^2)(\varepsilon^2 +\bp_+^2+\bDelta^2 )} 
\, ,
\label{eq:6.7}
\eea
where 
${\partial G(x,\bDelta^2)/ \partial \log \bDelta^2}={\cal F}(x,\bDelta^2)$ 
is the unintegrated
gluon structure function (SF) of the free nucleon \cite{INDiffGlue}.
\begin{figure}[t]
  \vspace{6.0cm}
  \includegraphics{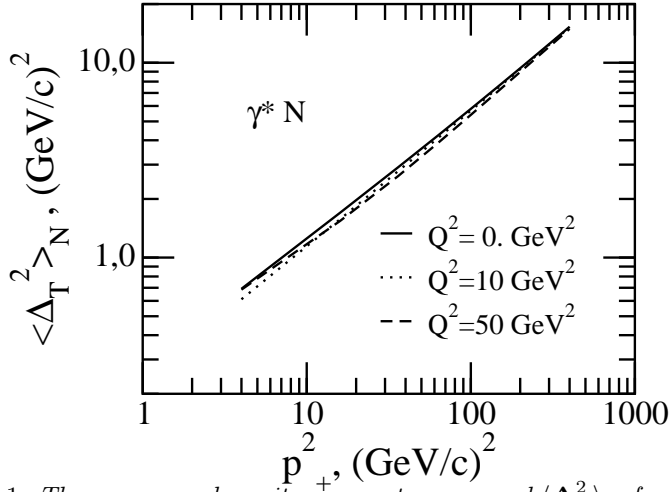}
  \caption{\it
The mean acoplanarity momentum squared 
$\langle \bDelta _{T}^2\rangle_N$ for DIS on a free nucleon
target with production of trigger jets with the transverse
momentum higher than $\bp_{+}$ for several values of $Q^{2}$.
 The 
numerical results are for $x=0.01$ and the input
unintegrated gluon SF of the proton is taken from ref.
\protect{\cite{INDiffGlue}}.
}
\label{exfig} 
\end{figure}
  Then a quick estimate
\bea
\langle \bDelta^2_{T} \rangle_N  
 \approx
\bp_+^2{{\cal F}(x,\bp_+^2) \over 2G(x,\bp_+^2)} \, ,
\label{eq:6.10}
\eea
correctly describes the numerical results 
shown in fig.~1. As far as the dijets are hard, 
$\bp_{+}^2 \gsim z(1-z)Q^2 \sim {1\over 4}Q^2$, the acoplanarity
momentum distribution would not depend on $Q^2$, which 
holds still better if one considers $\sigma_T +\sigma_L$.
This point is illustrated in fig.~1, where we show 
$\langle \bDelta^2_{T} \rangle_N$ at $z=1/2$ for several values of
$Q^2$. Because of this weak dependence on $Q^2$ here-below 
we make no distinction between DIS and real photoproduction, $Q^2=0$.
In the practical evaluations of the nuclear contribution
$\langle \bkappa_{T}^2(\bb) \rangle_A$ 
one can use the eq.(\ref{eq:4.10}) 
which gives the result                                           
\bea
\langle \bkappa_{T}^2 (\bb)\rangle_A 
\approx
{1\over 2}\lambda_cQ_A^2(\bb)
 \left[\log  {2p_+ 
\over \sqrt{\lambda_c}Q_A(\bb)}-1\right],
\label{eq:6.5} 
\eea 
where  
\bea
Q_A^2(\bb) \approx
{ 4\pi^2 \over N_c} \alpha_S(Q^2) G(x,Q^2) T(\bb)\, .
\label{eq:C.12}
\eea

\begin{figure}[t]
  \vspace{6.0cm}
  \includegraphics{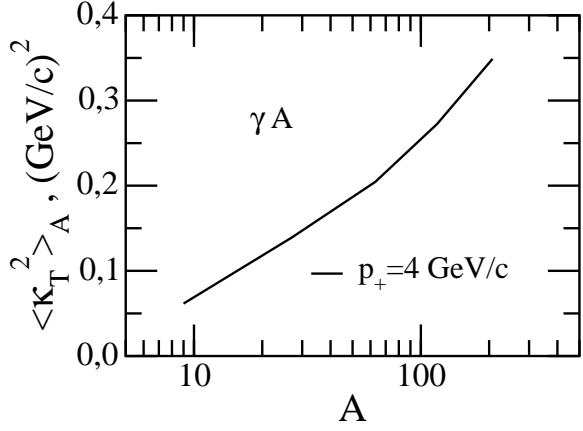}
  \caption{\it
The atomic mass number
dependence of nuclear broadening contribution, 
$\langle \bkappa _{T}^2(\bb)\rangle_A$,
 to the
mean acoplanarity momentum squared for real
photoproduction
off nuclei  at $x=0.01$. The input
unintegrated gluon SF of the proton is taken from ref.
\protect{\cite{INDiffGlue}.}
}
\label{exfig} 
\end{figure}
For average DIS on heavy nuclei the reference value is 
$\langle Q_{Au}^2(\bb)\rangle =0.9$ $(GeV/c)^2$,
see \cite{NSZZ}. The atomic mass number dependence of 
nuclear broadening 
$\langle \bkappa_{T}^2 \rangle_A$ for jets with  
$p_+ = 4$ GeV/c in average DIS off nucleus
is shown in fig.~2. The principal reason why 
$\langle \bkappa_{T}^2 \rangle_A$ is numerically
small compared to $\langle Q_{Au}^2(\bb)\rangle$ is
that even for such a heavy nucleus as $^{197}Au$ 
the no-broadening probability in average DIS is large,
$\langle w_{Au}(\bb,0)\rangle \approx 0.5$.  
A comparison of the free nucleon broadening $\langle \bDelta_{T}^2 
\rangle_N$ from fig.~1 with the nuclear
contribution $\langle \bkappa_{T}^2 (\bb)
\rangle_A$ from fig.~2  shows  that 
the nuclear mass number dependence of azimuthal decorrelation of 
dijets in average DIS off nuclei will be relatively weak. 

\begin{figure}[t]
  \vspace{6.0cm}
  \includegraphics{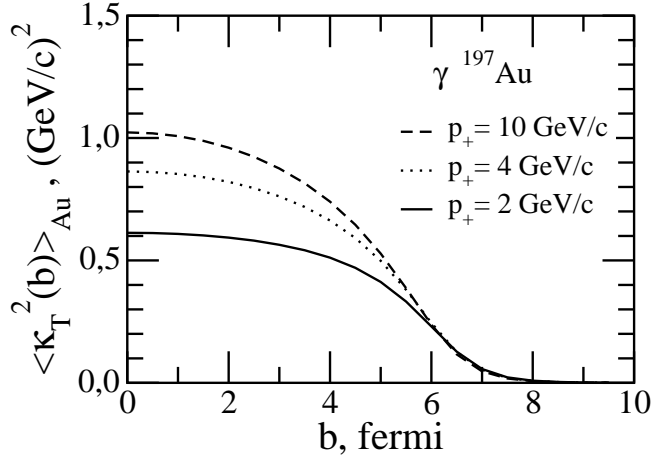}
  \caption{\it
 The impact parameter dependence of the 
nuclear broadening contribution, 
$\langle \bkappa _{T}^2(\bb)\rangle_A$,
 to the
mean acoplanarity momentum squared from 
peripheral DIS at large impact parameter 
to central DIS at $\bb=0$ for several
values of the away jet momentum $p_+$. The 
numerical results are for $x=0.01$ 
and the input
unintegrated gluon SF of the proton is taken from ref.
\protect{\cite{INDiffGlue}.}
}
\label{exfig} 
\end{figure}

However, nuclear broadening will be  substantially stronger
for a subsample of central DIS events at $\bb \sim 0$.
In fig.~3
we show for the gold, $^{197}Au$, target 
a dependence of the  averaged nuclear broadening 
$\langle \bkappa_{T}^2 (\bb)\rangle_A$ on the
impact parameter at several
values of $p_+$.

One can enhance $Q_{A}^2$ and nuclear contribution
$\langle \bkappa_{T}^2 (\bb)\rangle_A$
still further selecting DIS events when
the photon breaks up into the $q\bar{q}$ pair on the front face 
of a nucleus.
Experimentally, precisely such events are isolated by
selecting very large multiplicity or very high transverse 
energy of produced secondary particles
(\cite{RHIC_STAR} and references therein).

 Now we comment 
on the recent finding by the STAR collaboration 
of a disappearance of back-to-back high $p_{T}$ hadron 
correlation when going from peripheral to central
gold-gold collisions at RHIC \cite{RHIC_STAR}.
The application
of color dipole formalism to hard hadron-nucleus interactions
\cite{NPZcharm} suggests that our analysis of
acoplanarity of forward hard jets can be readily
generalized to mid-rapidity jets. One only has to choose an
appropriate system of dipoles, for instance, the open heavy
flavor production can be treated in terms of the
intranuclear propagation of the gluon-quark-antiquark system
in the overall color-singlet state. At RHIC energies jets
with moderately large $p_{T}$ are for the most part
due to gluon-gluon collisions. In our
language that can
be treated as a breakup of gluons into dijets and azimuthal
decorrelation of hard jets must be discussed in terms of
intranuclear propagation of color-octet gluon-gluon 
dipoles. For such gluon-gluon dipoles the relevant saturation scale
$Q_{8A}^2$ is larger \cite{NZ94} than that for the 
quark-antiquark dipoles
by the factor $2\lambda_c = C_A/C_F = 9/4$. Arguably, in
central nucleus-nucleus collisions distortions in the
target and projectile nuclei add up and the effective thickness
of nuclear matter is about twice of that in DIS. 
Then,
the results shown in fig.~2 suggest that
for central gold-gold collisions the nuclear broadening 
of gluon-gluon dijets
could be quite substantial, 
$\langle \bkappa_{T}^2(\bb=0)\rangle_{AuAu} \sim $(3-4) $(GeV/c)^2$
for average central $AuAu$ collisions and even twice larger if
collisions take place at front surface of colliding nuclei. 
 
The 
principal effect of nuclear broadening is a reduction of the probability
of observing the back-to-back jets 
\beq
P(b)\approx{\langle \bDelta_{T}^2
\rangle_N \over  
\langle \bkappa_{T}^2(\bb) 
\rangle_A + \langle \bDelta_{T}^2
\rangle_N}
\label{eq:6.11}
\eeq
and one needs to compare  
$\langle \bDelta_{T}^2 
\rangle_N$ to $\langle \bkappa_{T}^2 (\bb)
\rangle_A \, .
$
Our eq.~(\ref{eq:6.10}) for the free 
nucleon case holds as well for the gluon-gluon
collisions. Then the results shown in
fig.~1 entail  $\langle \bDelta_{T}^2 
\rangle_N \approx \langle \bkappa_{T}^2(0)\rangle_{AuAu} 
\sim $(3-4) $(GeV/c)^2$ at the jet momentum $p_+ = p_J =$ (6-8) GeV/c
and our nuclear
broadening 
would become substantial for all jets with $p_+$ below
the decorrelation
threshold momentum $p_J$.
In practice, the STAR  collaboration studied the azimuthal correlation of 
two high-$p_{T}$ hadrons and for the quantitative correspondence
between the STAR observable and azimuthal decorrelation in the parent 
dijet one needs to model fragmentation of jets into hadrons (for the 
modern fragmentation schemes see \cite{LUND}), here we notice that
the cutoff $p_+$ in our
Gedanken experiment is related to the momentum 
cutoff $p_{T,min}$ of  associated tracks from the away jet, 
whereas our jet of
momentum $\bp_-$ can be regarded as a counterpart of the trigger
jet of STAR. The STAR cutoff $p_T = $ 2GeV/c corresponds to the 
parents jets with the transverse momentum $p_+\sim (2-3)p_T
=$(4-6) GeV which is 
comparable to, or even smaller than, the decorrelation
threshold momentum
 $p_J= $ (6-8) GeV/c. Then eq. (\ref{eq:6.11}) suggests that 
in the kinematics of STAR the probability to observe the
back-to-back away and trigger jets decreases 
approximately twofold from peripheral to central $Au Au$
collisions, 
$$
P(0)\approx 0.5\,,
$$
and 
perhaps even stronger, so that  
our azimuthal decorrelation may contribute substantially
to the STAR effect.

\section{Remark on breakup of photons into semihard dijets}

In our previous analysis \cite{Saturation} of the single 
particle spectrum it has been discovered
that the transverse momentum distribution of sea quarks
is dominated by the anticollinear, anti-DGLAP splitting of gluons
into sea, when the transverse momentum of the parent gluons is larger
than the momentum of the sea quarks. 
That suggests strongly
a complete azimuthal decorrelation of forward minijets with the transverse
momenta below the saturation scale, $p_{\pm} \lsim Q_A$.
In \cite{NSZZ} this limiting case has been  considered in detail.
The principal point is that the minijet-minijet inclusive cross
section depends
on neither the minijet nor decorrelation momentum. This observation proves
a disappearance of the azimuthal correlation
of minijets with the transverse momentum below the saturation scale.

\section{Acknowledgments}
It is a pleasure to thank Mario Greco for inviting me to talk. Special 
thanks to all 
organizers for making the Conference run so smoothly.

\end{document}